\documentstyle[psfig]{article}
\begin{document}
\begin{center}
   {\huge \bf The first three years of the outburst and light-echo 
          evolution of V838 Mon and the nature~of its progenitor}

\vskip 0.5 cm
{\Large U. Munari$^1$ and A.Henden$^2$}\\
\vskip 0.3 cm
$^1$ INAF National Institute of Astrophysics, I-36012 Asiago (VI), Italy\\
$^2$ Univ. Space Research Assoc./U. S. Naval Obs., P. O. Box 1149, Flagstaff AZ 86002-1149, USA\\
\end{center}

\vskip 0.5 cm
\noindent
{\footnotesize {\bf Abstract}.
V838~Mon has undergone one of the most mysterious stellar outbursts on
record, with ($a$) a large amplitude ($\Delta B$$\sim$10~mag) and
multi-maxima photometric pattern, ($b$) a cool spectral type at maximum
becoming cooler and cooler with time during the descent, until it reached
the never-seen-before realm of L-type supergiants, never passing through
optically thin or nebular stages, ($c$) the development of a spectacular,
monotonically expanding light-echo in the circumstellar material, and ($d$)
the identification of a massive and young B3V companion, unaffected by the
outburst. In this talk we review the photometric and spectroscopic evolution
during the first three full years of outburst, the light-echo development
and infer the nature of the progenitor, which was brighter and hotter in
quiescence than the B3V companion and with an inferred ZAMS mass of
$\sim$65~M$_\odot$.}\\

\vskip 0.5 cm
\centerline{\Large \bf The outburst evolution}

\vskip 0.5 cm 
\noindent
V838~Mon made headlines in early January 2002, when it was discovered in
outburst by [1]. The unusually cool spectrum (completely unlike
that of a classical nova) and the multi-maxima light-curve helped to keep
attention focused on the object for the next three months, until the
discovery in late March by [2] of a light-echo rapidly developing
around V838~Mon. The presence of the first Galactic light echo in $\sim$70
years fostered a massive, multi-wavelength observing campaign for V838~Mon.
A high spatial resolution imaging series of the light-echo expansion and
evolution was collected with HST by [3], which has been expanded by
new images secured within the Hubble Heritage
Program\footnote{http://heritage.stsci.edu}. An account of the
spectroscopic, photometric and polarimetric evolution of V838~Mon during the
first season of visibility was presented by [4], [5], [6], [7], [8], [9], [10], [11].
A major observational constraint was the
discovery by [12] and [13] that V838~Mon is a binary
system containing a normal B3V star, implying that the outbursting component
is young and massive. BaII, LiI and $s-$element lines were prominent in the
outburst spectra, while Balmer lines emerged with modest emission only
during the central phase of the outburst. The outburst never reached
optically thin conditions and the spectra never went through a nebular
stage.

An updated optical and infrared lightcurve of the eruption of V838~Mon is
presented in Figure~1 and the general spectral evolution is highlighted in
Figure~2. The outburst lightcurve and spectral evolution can be easily
divided into three distinctive phases: the K, M and L supergiant spectrum
phases.

\clearpage
\begin{figure}[!H]
\centerline{\psfig{file=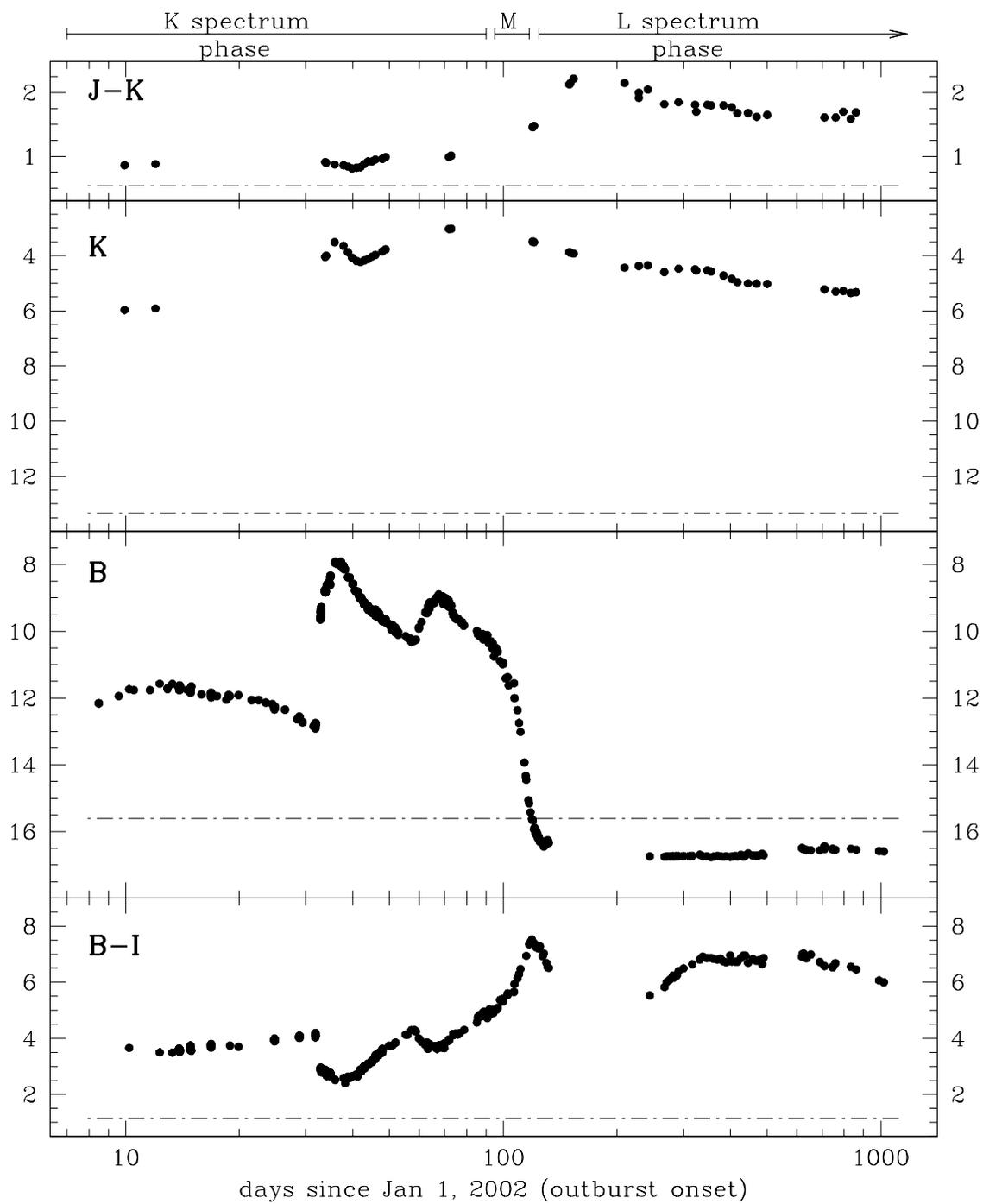,width=15.0cm}}
      \caption{Optical and infrared lightcurves of V838~Mon from 2002-2004
               observations with the USNO 1.0 and 1.55m telescopes.}
\end{figure}

\begin{figure}[!H]
\centerline{\psfig{file=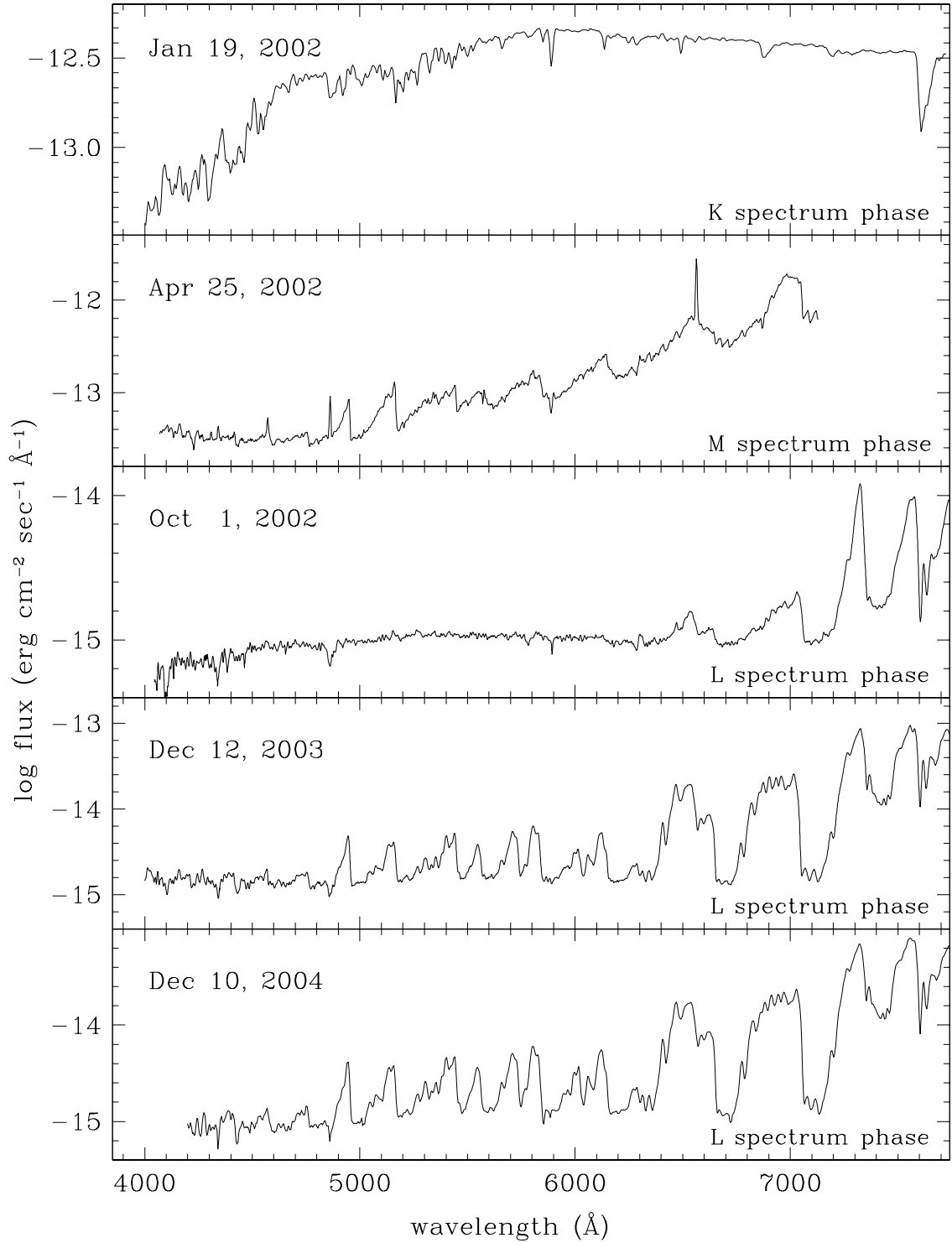,width=15.0cm}}
      \caption{Spectral evolution of V838~Mon (sample spectra from an
               extensive monitoring of the whole event performed with the
               1.82 m Asiago telescope, to be reported in detail by Munari
               et al. elsewhere).}
\end{figure}
\clearpage

\begin{figure}[!t]
\centerline{\psfig{file=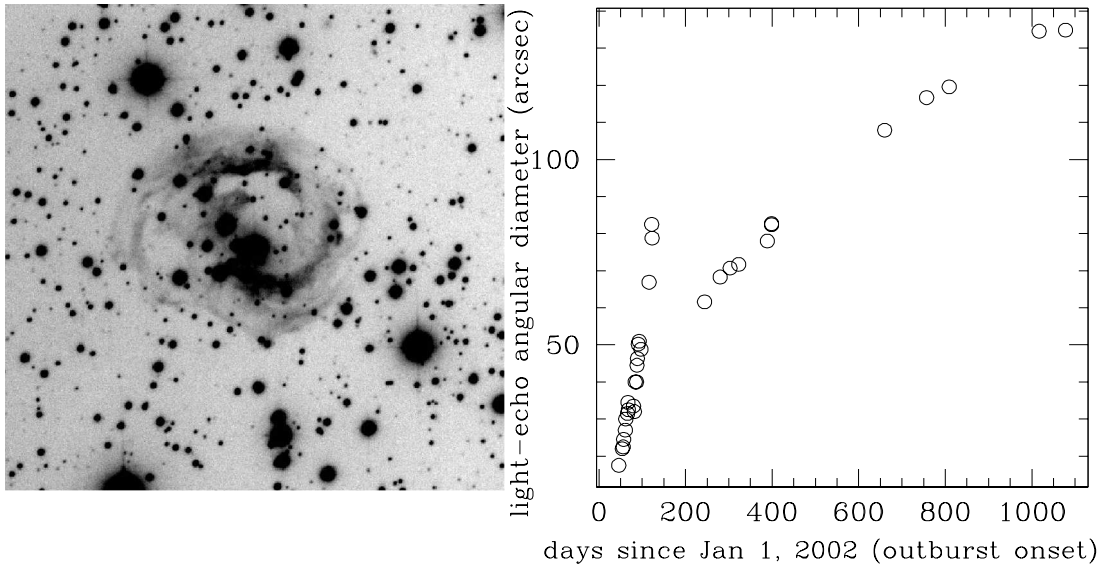,width=14.5cm}}
      \caption{{\em Left panel}. The light echo imaged in $R_{\rm C}$ band
               with the 1.55 m USNO telescope on Oct 15, 2004 (48 min total
               exposure). The longest dimension of the light-echo occurs at
               position angle 100$^\circ$ where it extends 2.37 arcmin.
               Field is 4.5 arcmin square. North is up, East to the left. 
               {\em Right panel}. The expansion rate of the outer edge of
               the light echo in the east-west direction. Based on NOFS
               ground-based imagery in multiple bandpasses.}

\end{figure}

\begin{figure}[!t]
\centerline{\psfig{file=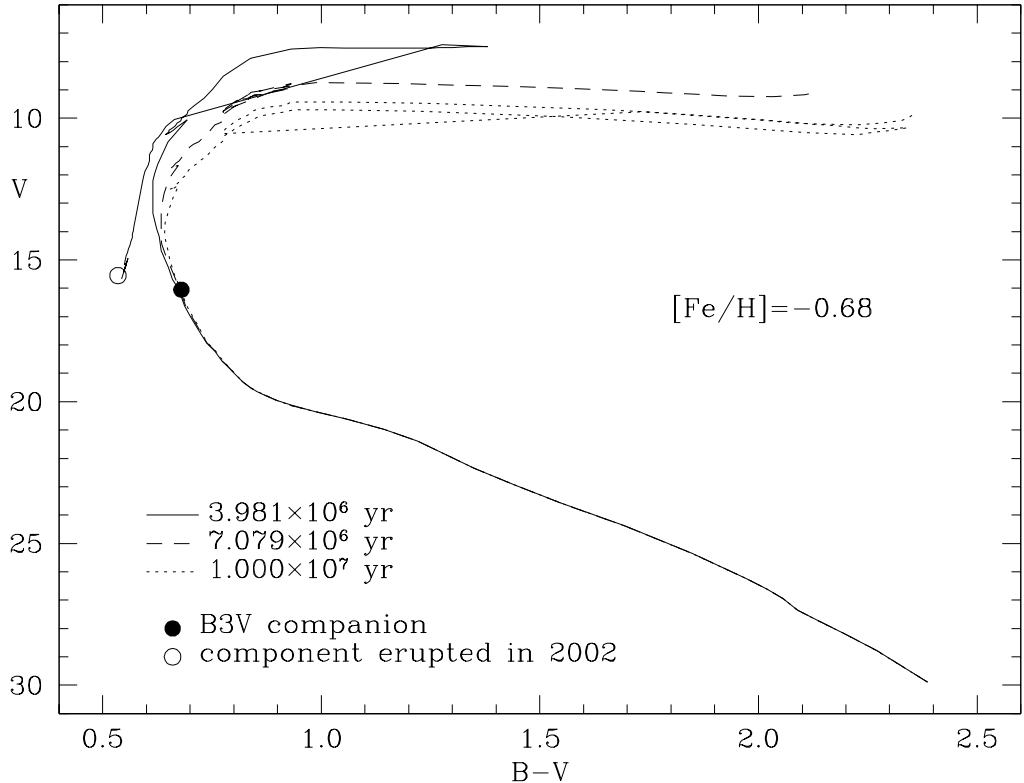,width=13.5cm}}
       \caption{Comparison of the pre-outburst photometry of the two
               components of V838~Mon with Padova isochrones Z=0.004 and
               three ages scaled to the system distance (10 kpc) and
               reddening ($E_{\rm B-V}$=0.87). The isochrones include the
               effect of mass loss and reddening-induced distortion.}
\end{figure}

During the first 90 days of the outburst, the star went through three maxima
and its spectrum moved back and forth within the K supergiant types. The
monotonic cooling during this phase (from $B$$-$$I_{\rm C}$$\sim$3.5 to
$\sim$5.0, cf. Figure~1) was interrupted by warming at each of three
optical brightness maxima. This {\em K-supergiant} phase was characterized
by a cool wind that gave rise to P-Cyg profiles with terminal velocities of
$\sim$500~km~sec$^{-1}$ that progressively diminished to
$\sim$250~km~sec$^{-1}$ by day +90. 

Around day 90, the star entered an {\em M-supergiant} phase almost identical
to that displayed by the M31-RV stellar eruption observed in 1988 in the
Andromeda Galaxy as described by [14] and [15]. [16]
investigated in detail the similarities, highlighting how in both events the
transition to an M-supergiant spectrum was marked by a free-fall drop in
optical brightness not accompanied by dust formation. The outbursting
component in a matter of few weeks swept through the whole sequence of M
spectral types while the $B$ band brightness of V838~Mon dropped by 6~mag,
stopped only by the emergence of the $B$=16.73~mag spectrum of the B3V
companion by day $\sim$120. 

By day $\sim$130, the photosphere of the outbursting component had cooled so
much to enter the realm of {\em L supergiant} spectral types, a type of star
never seen before anywhere in the Galaxy ([12], [17]).
Very strong H$_2$O, CO and AlO absorption bands dominate the infrared, while
huge VO and TiO shape the optical. Since the time of coolest temperature
reached around day 180, the temperature of the L-supergiant has been
smoothly and constantly increasing even if by small amounts (cf. $J$$-$$K$
and $B$$-$$I$ in Figure~2). As a result, the optical spectra that in Oct
2002 were dominated by the B3V companion even in the $V$ band, are now again
showing the molecular absorptions down to $\sim$4800~\AA.

\vskip 0.5 cm
\centerline{\Large \bf The light-echo evolution}

\vskip 0.5 cm
\noindent
The light-echo was first discovered by [2] on U-band images taken
in February, 2002, about 10 days after the first large outburst (a
hundred-fold increase in brightness with a relatively sharp peak).  The
initial shell was circularly symmetric, showing obvious structure that later
HST images by [3] resolved into concentric rings, each giving a
photometric snapshot of the light curve variations.  These rings expanded,
with the rate of expansion of the outermost ring shown in Figure~3.  Note
the dramatic break in the expansion rate, with the early expansion
proceeding at about 800mas per day, while the later expansion proceeding at
about 90mas per day.  The early expansion is probably coming from a light
sheet in front of the outbursting star, and is very nearly symmetric, with
axial ratios at most 10\% different.  The later expansion may be
evidence of the light echo proceeding through circumstellar material, and
shows more than 20\% asymmetry.  The light-echo has also faded, now
difficult to image even at R-band, and has become more extended in the
east-west direction than north-south. An inner void appeared about 70 days
after the main outburst, matching the rapid decline from the last
photometric peak. The bright cloud to the north has remained in all images.
More detailed analysis of the ring expansion and structure evolution is in
progress.

\vskip 0.5 cm
\centerline{\Large \bf The progenitor}

\vskip 0.5 cm
\noindent
The energy distribution of V838~Mon in quiescence has been throughly
investigated by [18], clearly showing how {\em the progenitor of
the outbursting component was brighter and hotter than the B3\,V companion}.
The best fit to the V838~Mon quiescence data is the B3\,V plus a 50\,000~K
star with $V$=15.55 and $B-V$=+0.535, both equally reddened by
$E_{B-V}$=0.87. The location of the components of the binary is compared
with Padova theoretical isochrones in Figure~4 (scaled to the system
$E_{B-V}$=0.87 reddening and 10~kpc distance), for the [Fe/H]=$-$0.7
metallicity appropriate to the galacto-centric distance of V838~Mon. The
Padova theoretical isochrones (that include the effect of mass loss) are
corrected for reddening dependent deformation following [19].

The isochrone fitting the position of the two components indicates a system
age of 4 million yr and a mass of $\sim$65~M$_\odot$ on the ZAMS for the
progenitor of the outbursting component. The outburst experienced in 2002
does not appear to be the terminal event in the life of the massive progenitor,
but instead more probably was a thermonuclear shell flash in the outer layers
of the star as could be expected in the case of He after most of the H-rich
outer envelope has been blown away by the strong wind that characterize the
Wolf-Rayet type of stars that occupy this part of the HR diagram.

\vskip 0.5 cm
\centerline{\Large \bf References}

\footnotesize

\vskip 0.5 cm
\noindent
$[$1$]$ Brown N.J. 2002, IAU Circ 7785\\
$[$2$]$ Henden, A., Munari, U., Schwartz, M.B., 2002, IAUC 7859 \\
$[$3$]$ Bond, H.E., Panagia, N., Sparks, W.B., Starrfield, S.G.,
     Wagner, R.M., 2002, IAUC 7892\\
$[$4$]$ Munari, U., Henden, A., Kiyota, S., et al., 
     2002, A\&ALett 389, L51\\
$[$5$]$ Banerjee, D.P.K., Ashok, N.M. 2002, A\&A 395, 161\\
$[$6$]$ Goranskii, V.P., Kusakin, A.V., Metlova, N.V. et al. 
     2002, Astron.Lett. 28, 691\\
$[$7$]$ Kimeswenger, S., Lederle, C., Schmeja, S. et al. 
     2002, MNRAS 336, L43\\
$[$8$]$ Crause, L.A., Lawson, W.A., Kilkenny, D. et al. 
     2003, MNRAS 341, 785\\
$[$9$]$ Wisniewski, J.P., Morrison, N.D., Bjorkman, K.S. et al. 
     2003, ApJ 588, 486\\
$[$10$]$ Desidera, S., Giro, E., Munari, U. et al. 2004, A\&A 414, 591\\
$[$11$]$ Kipper, T., Klochkova, V.G., Annuk, K. et al. 
      2004, A\&A 416, 1107\\
$[$12$]$ Desidera, S., Munari, U. 2002, IAUC 7982\\
$[$13$]$ Munari, U., Desidera, S., Henden, A., 2002, IAUC 8005\\
$[$14$]$ Rich, R.M., Mould, J., Picard, A., et al., 1989, ApJ 341, L51 \\
$[$15$]$ Mould, J., Cohen, J., Graham, J.R., et al., 
      1990, ApJ 353, L35 \\
$[$16$]$ Boschi F., Munari U. 2004, A\&A 418, 869\\
$[$17$]$ Evans A., Geballe T.R., Rushton M.T. et al. 
      2003, MNRAS 343 1054\\
$[$18$]$ Munari, U., Henden, A., Vallenari, A. et al. 
      2004, A\&A, submitted July 29, 2004\\
$[$19$]$ Fiorucci, M., \& Munari, U. 2003, A\&A, 401, 781\\

\end{document}